\documentclass[prd,preprint,superscriptaddress,showpacs,byrevtex]{revtex4}
\usepackage{bm}
\def\fsl#1{\setbox0=\hbox{$#1$}           
   \dimen0=\wd0                                 
   \setbox1=\hbox{/} \dimen1=\wd1               
   \ifdim\dimen0>\dimen1                        
      \rlap{\hbox to \dimen0{\hfil/\hfil}}      
      #1                                        
   \else                                        
      \rlap{\hbox to \dimen1{\hfil$#1$\hfil}}   
      /                                         
   \fi}                                         %
\newcommand{\be}{\begin{equation}}
\newcommand{\ee}{\end{equation}}
\newcommand{\bea}{\begin{eqnarray}}
\newcommand{\eea}{\end{eqnarray}}
\newcommand{\beq}{\begin{equation}}
\newcommand{\eeq}{\end{equation}}
\newcommand{\beqs}{\begin{eqnarray}}
\newcommand{\eeqs}{\end{eqnarray}}

\newcommand{\dslash}{D\hspace{-0.067in}\slash}

\newcommand{\qslash}{Q\hspace{-0.067in}\slash}

\newcommand{\aslash}{A\hspace{-0.067in}\slash}
\newcommand{\delslash}{\partial\hspace{-0.067in}\slash}
\begin{document}
\title{ Factorization of Soft and Collinear Divergences in Non-Equilibrium Quantum Field Theory }
\author{Gouranga C Nayak} \email{nayak@physics.arizona.edu}
\affiliation{Department of Physics, University of Arizona, Tucson 85721, USA }
\begin{abstract}
Proof of factorization of soft and collinear divergences in non-equilibrium QCD may be
necessary to study hadronic signatures of quark-gluon plasma at RHIC and LHC.
In this paper we prove factorization of soft and collinear divergences in non-equilibrium
QED by using Schwinger-Keldysh closed-time path integral formalism in the
background field method in pure gauge.
\end{abstract}
\pacs{ PACS: 13.87.Fh, 11.10.-z, 11.10.Wx, 11.15.Kc }
\maketitle
\pagestyle{plain}
\pagenumbering{arabic}
\section{Introduction}
RHIC and LHC heavy-ion colliders are the best facilities to study quark-gluon
plasma in the laboratory. Since two nuclei travel almost at speed of light,
the QCD matter formed at RHIC and LHC may be in non-equilibrium. In order
to make meaningful comparison of the theory with the experimental data on hadron
production, it may be necessary to study nonequilibrium-nonperturbative QCD at
RHIC and LHC. This, however, is a difficult problem.

Non-equilibrium quantum field theory can be studied by using Schwinger-Keldysh closed-time
path (CTP) formalism \cite{schw,keldysh}. However, implementing CTP in non-equilibrium at
RHIC and LHC is a very difficult problem, especially due to the presence of gluons in
non-equilibrium and hadronization etc. Recently, one-loop resummed gluon propagator in
non-equilibrium in covariant gauge is derived in \cite{greiner,cooper}.

High $p_T$ hadron production at high energy $e^+e^-$, $ep$ and $pp$ colliders is studied
by using Collins-Soper fragmentation function \cite{collins,sterman,george}. For a high $p_T$
parton fragmenting to hadron, Collins-Soper derived an expression for the fragmentation
function based on field theory and factorization properties in QCD at high energy. This
fragmentation function is universal in the sense that, once its value is determined from
one experiment it explains the data at other experiments \cite{frag}.

Recently we have derived parton-to-hadron fragmentation function in non-equilibrium QCD
by using Schwinger-Keldysh closed-time path integral formalism \cite{nayakfrag}.
This can be relevant at RHIC and LHC heavy-ion colliders to study hadron production
from quark-gluon plasma. We have considered a high $p_T$ parton in QCD medium at
initial time $\tau_0$ with arbitrary non-equilibrium (non-isotropic) distribution function
$f(\vec{p})$, fragmenting to hadron. The special case $f(\vec{p})=\frac{1}{e^{\frac{p_0}{T}}\pm 1}$
corresponds to the finite temperature QCD in equilibrium.

In order to study hadron production from quark-gluon plasma using the non-equilibrium fragmentation
function we will need to prove factorization of fragmentation function in non-equilibrium QCD. Factorization refers to separation
of short distance from long distance effects in field theory. In this paper we will prove the factorization of
soft and collinear divergences in non-equilibrium QED. The proof of factorization of fragmentation function
in non-equilibrium QCD at RHIC and LHC will be addressed elsewhere.

The paper is organized as follows. In section II we briefly describe factorization of soft and
collinear divergences in QED in vacuum. In section III we describe Schwinger-Keldysh closed-time path integral formalism in non-equilibrium
quantum field theory relevant for our purpose. In section IV we prove factorization of soft and collinear divergences
in non-equilibrium QED by using Schwinger-Keldysh closed-time path formalism in the background field method in pure gauge.
Section V contains conclusions.

\section{ Factorization of Soft and Collinear Divergences in QED in Vacuum }

The Wilson line is given by \cite{sterman,tucci}
\bea
\Phi[x^\mu ]={\cal P}~ {\rm exp}[-ie\int_{-\infty}^{0} d\lambda~ h \cdot A(x^\mu +h^\mu \lambda )]
\label{wilf}
\eea
where $h^\mu$ is a $x^\mu$ independent four vector. Eq. (\ref{wilf}) can be written as
\bea
&&~\Phi[x^\mu ]={\cal P}~ {\rm exp}[-ie\int_{-\infty}^0 d\lambda~ h \cdot A(x^\mu +h^\mu \lambda )] \nonumber \\
&& ={\cal P}~ {\rm exp}[-ie\int_{-\infty}^0 d\lambda~ h \cdot e^{\lambda h \cdot \partial}  A(x^\mu )]
={\cal P}~ {\rm exp}[-ie \frac{1}{ h \cdot \partial} h \cdot  A(x^\mu )].
\label{wilf1}
\eea
Since the vector $h^\mu$ is free we can choose it to correspond to various physical situations.
For example if we choose $h^\mu=n^\mu$, where $n^\mu$ is a fixed lightlike vector, we find from eq. (\ref{wilf1}) the phase factor
\bea
\omega(x)= \frac{1}{ n \cdot \partial} n \cdot  A(x ).
\label{eik}
\eea
Using the Fourier transformation
\bea
A_\mu(x) =\int \frac{d^4k}{(2\pi)^4} A_\mu(k) e^{ik \cdot x}
\label{ft}
\eea
we find from eq. (\ref{eik})
\bea
V=e~\omega(k)=ie~\frac{n^\mu}{n \cdot k} A_\mu(k).
\label{eikonal}
\eea
Note that eq. (\ref{eikonal}) is precisely the eikonal vertex for a soft photon with momentum $k$ interacting with a
high energy electron jet moving along the direction $n^\mu$ \cite{collins,sterman}. In the soft photon
approximation in \cite{sterman} the fixed lightlike vector is taken to be having only "+" or "-" component:
\bea
n^\mu=(n^+,n^-,n_T)=(1,0,0)~~~~~~~~~~~{\rm or}~~~~~~~~~n^\mu=(n^+,n^-,n_T)=(0,1,0).
\label{n}
\eea

Now we will show that when the classical background field $A_\mu(x)$ is a pure gauge given by
\bea
A_\mu(x) = \partial_\mu \omega(x)
\label{om1}
\eea
we can reproduce eqs. (\ref{eik}) and (\ref{eikonal}) which appear in the Wilson line eq. (\ref{wilf1}).
Now multiplying a four vector $h^\mu$ from left in eq. (\ref{om1}) we find
\bea
h \cdot A(x) = h \cdot \partial \omega(x).
\label{om2}
\eea
Dividing $h \cdot \partial $ from left in the above equation we find
\bea
\omega(x) = \frac{1}{h \cdot \partial } h \cdot A(x).
\label{om3}
\eea
Since the four vector $h^\mu$ is free we can choose it such way that the pure gauge can represent Feynman rules
involving soft photons or collinear photons. For example, when $h^\mu =n^\mu$ where $n^\mu$ is a fixed vector (see
eq. (\ref{n})), we find from the above equation
\bea
\omega(x) = \frac{1}{n \cdot \partial } n \cdot A(x)
\label{om4}
\eea
which reproduces eq. (\ref{eik}) which appears inside the Wilson line in eq. (\ref{wilf1}). Hence we have
established the correspondence between the Wilson line and the classical background field $A_\mu(x)$ in
pure gauge in the context of soft divergences.

Similarly, if we choose $h^\mu=n^\mu_B$, where $n^\mu_B$ is a non-light like vector
\bea
n^\mu_B=(n^+_B,n^-_B,0),
\label{nb}
\eea
we reproduce the Feynman rules for the collinear divergences \cite{sterman}.
This establishes the correspondence between the Wilson line and the classical background
field $A_\mu(x)$ in pure gauge in the context of collinear divergences.

The generating functional in QED in the presence of background field $A_\mu(x)$ is given by \cite{tucci}
\bea
Z[A,J,\eta,{\bar \eta}]=\int [dQ] [d{\bar \psi}] [d \psi ]~
e^{i\int d^4x [-\frac{1}{4}{F}_{\mu \nu}^2[Q] -\frac{1}{2 \alpha} (\partial_\mu Q^{\mu })^2+{\bar \psi} \dslash [A+Q] \psi
+ J \cdot Q +{\bar \eta} \psi + \eta  {\bar \psi} ]}
\label{zfqi}
\eea
where
\bea
\dslash [A+Q] = (i \delslash - e\qslash -e \aslash ).
\label{dsl}
\eea
For the purpose of studying soft and collinear divergences it is understood that the integration in
$Z[A,J,\eta,{\bar \eta}]$ is only over those functions of the quantum photon field $Q(x)$, electron field
$\psi(x)$ and positron field ${\bar \psi}(x)$ such that their Fourier transforms $Q(k)$, $\psi(k)$ and ${\bar \psi}(k)$
vanish in the soft region. Under the following transformation of the fermion fields and the sources
\bea
\psi' = U~ \psi,~~~~{\bar \psi}' = {\bar \psi} ~U^{-1},~~~~~~\eta' = U~ \eta,~~~~{\bar \eta}' = {\bar \eta}~ U^{-1},~~~~~~~~U=e^{-ie\omega(x)}
\label{ft}
\eea
we find from eq. (\ref{zfqi})
\bea
Z[A,J,\eta',{\bar \eta}']~=~Z[J,\eta,{\bar \eta}].
\label{zz}
\eea

The correlation function in QED in the presence of background field $A_\mu(x)$ is given by
\bea
\frac{\delta}{\delta {\bar \eta}(x_2) }~\frac{\delta}{\delta { \eta}(x_1) }~Z[A,J,\eta,{\bar \eta}]~|_{J=\eta={\bar \eta}=0}=
<\psi(x_2) {\bar \psi}(x_1)>_A.
\label{g1}
\eea
Similarly the correlation function in QED (without the background field) is given by
\bea
\frac{\delta}{\delta {\bar \eta}(x_2) }~\frac{\delta}{\delta { \eta}(x_1) }~Z[J,\eta,{\bar \eta}]~|_{J=\eta={\bar \eta}=0}=
<\psi(x_2) {\bar \psi}(x_1)>_{A=0}.
\label{g2}
\eea

Hence we find from eqs. (\ref{zz}), (\ref{g1}) and (\ref{g2})
\bea
<\psi(x_2) {\bar \psi}(x_1)>_A~=~e^{-ie \omega(x_2)}~<\psi(x_2) {\bar \psi}(x_1)>_{A=0}~e^{ie \omega(x_1)}
\label{fact1q}
\eea
where all the $A_\mu(x)$ dependence has been factored into $e^{-ie \omega(x_2)}$ and $e^{ie \omega(x_1)}$.
Using eqs. (\ref{om3}) and (\ref{wilf1}) we find
\bea
&&~<\psi(x_2) {\bar \psi}(x_1)>_A~=~
{\rm exp}[-ie\int_{-\infty}^0 d\lambda~ h \cdot A(x_2^\mu +h^\mu \lambda )]~\nonumber \\
&& ~\times ~<\psi(x_2) {\bar \psi}(x_1)>_{A=0}~\times ~{\rm exp}[-ie\int_0^{-\infty} d\lambda~ h \cdot A(x_1^\nu +h^\nu \lambda )].
\label{fact2t}
\eea
This proves factorization of soft and collinear divergences in QED \cite{tucci}.
\section{ Schwinger-Keldysh Closed-Time Path Integral Formalism in Non-Equilibrium Quantum Field Theory }
Unlike $pp$ collisions, the ground state at RHIC and LHC heavy-ion collisions
(due to the presence of a QCD medium at initial time $t=t_{in}$ (say $t_{in}$=0)
is not a vacuum state $|0>$ any more. We denote $|in>$ as the initial state of the non-equilibrium QCD
medium at $t_{in}$. For example, if the system at RHIC and LHC at initial time is space
translational invariance the non-equilibrium
distribution function $f(\vec{k})$ of a parton (quark or gluon),
corresponding to such initial state can be written as
\bea
<a^\dagger ({\vec k})a({\vec k}')>=<in|a^\dagger ({\vec k})a({\vec k}')|in> = f(\vec{k}) (2\pi)^{d-1} \delta^{(d-1)} ({\vec k} -{\vec k}').
\label{dist}
\eea
Finite temperature field theory formulation is a special case of this when
$f({\vec k}) =\frac{1}{e^{\frac{k_0}{T}} \pm 1}$.

Consider the time evolution of the density matrix
\bea
i\frac{\partial \rho}{\partial t} = [H_I, \rho],~~~~~~~~~~~~~~~~\rho(-\infty) = \rho_0
\label{rho1}
\eea
where $H_I$ is the interaction hamiltonian. The formal solution is
\bea
\rho(t)=S(t,-\infty)\rho_0 S(-\infty,t),~~~~~~~~~~~~~~~~~~~S(t,-\infty)=T~exp[-i\int_{-\infty}^t dt'~H_I(t')].
\label{rho2}
\eea
The density matrix $\rho$ is in interaction picture. The average value of an operator $L$ in the interaction
picture is given by
\bea
<L(t)>=Tr[\rho(t)L(t)],~~~~~~~~~~~~~~~~~~~~~~~~~~i\frac{\partial L(t)}{\partial t} = [H_0, L(t)]
\label{rho3}
\eea
where $H_0$ is the free hamiltonian. Since in many situations we deal with correlation function of several fields
at different times, it is useful to transfer all the time dependence to field operators and consider the density
operator as independent of time, {\it i.e.} to go to Heisenberg representation. For the time independence of the density
matrix we can take the value of the matrix determined by expression eq. (\ref{rho2}) at a certain fixed instant
of time, for example, $t=t_{in}=0$, having thus included in it all the changes which the distribution $\rho_0$ had
undergone when the external field and the interaction in the system were switched on.

Consider the medium average of $T-$products of several operators. Using the Heisenberg density matrix one finds
\bea
&& <T[L(t)M(t')...]>=Tr[\rho T[L(t)M(t')....]]=Tr[ S(0,-\infty)\rho_0 S(-\infty,0)T[L(t)M(t')....]] \nonumber \\
&& =Tr[ \rho_0 S(-\infty,0)T[L(t)M(t')....]S(0,-\infty)].
\label{rho4}
\eea
Going over to the operators in the interaction picture
\bea
&&<T[L(t)M(t')...]>=Tr[ \rho_0 S(-\infty,0)T[S(0,t)L_I(t)S(t,t')M_I(t')....]S(0,-\infty)] \nonumber \\
&&=Tr[ \rho_0 T_c[S_cL_I(t)M_I(t')....]]
\label{rho5}
\eea
where $T_c$ is the complete contour from
\bea
-\infty \rightarrow t \rightarrow t' ....\rightarrow -\infty
\label{rho6}
\eea
and $S_c$ is the complete $S-$matrix defined along $T_c$.

To deal with Feynman diagram and Wick theorem it is useful to split the time interval to "+" and "-" contour;
where "+" time branch is from $-\infty$ to $+\infty$ where (time) $T-$order product apply and "-" time branch
is from $+\infty$ to $-\infty$ where (anti-time)${\bar T}-$order product apply.

\subsection{ Generating Functional in Non-Equilibrium Scalar Field Theory }
Consider scalar field theory first. Since there are two time
branches there are two fields and two sources and hence four Green's functions.
Let us denote the field $\phi_+(x)$ and the source $J_+(x)$ in the "+" time branch
and $\phi_-(x)$ and $J_-(x)$ in the "-" time branch. The generating functional is given by
\bea
Z[J_+,J_-,\rho]=\int [d\phi_+][d\phi_-] ~{\rm exp}[i[S[\phi_+]-S[\phi_-]+\int d^4x J_+\phi_+-\int d^4x J_-\phi_-]]~<\phi_+,0|\rho|0,\phi_-> \nonumber \\
\label{rho7}
\eea
where $S[\phi]$ is the full action in scaler field theory and $|\phi_{\pm},0>$ is the quantum state corresponding to the field
configuration $\phi_{\pm}(\vec{x},t=0)$.

In the CTP formalism in non-equilibrium there are four Green's functions
\bea
&& G_{++}(x,x') = \frac{\delta Z[J_+,J_-,\rho]}{i^2 \delta J_+(x) J_+(x')}=<in|T\phi (x) \phi (x')|in> = <T\phi (x) \phi (x')>\nonumber \\
&& G_{--}(x,x') = \frac{\delta Z[J_+,J_-,\rho]}{(-i)^2 \delta J_-(x) J_-(x')}= <in|{\bar T} \phi (x) \phi (x')|in> = <{\bar T} \phi (x) \phi (x')>\nonumber \\
&& G_{+-}(x,x') = \frac{\delta Z[J_+,J_-,\rho]}{-i^2 \delta J_+(x) J_-(x')}= <in|\phi (x') \phi (x)|in> = <\phi (x') \phi (x) >\nonumber \\
&& G_{-+}(x,x') = \frac{\delta Z[J_+,J_-,\rho]}{-i^2 \delta J_-(x) J_+(x')}= <in|\phi (x) \phi (x')|in>= <\phi (x) \phi (x') >
\label{green}
\eea
where $T$ is the time order product and ${\bar T}$ is the anti-time order product given by
\bea
&& T\phi (x) \phi (x') = \theta(t - t') \phi (x) \phi (x') + \theta(t'-t) \phi (x') \phi (x) \nonumber \\
&& {\bar T} \phi (x) \phi (x') = \theta(t' - t) \phi (x) \phi (x') + \theta(t-t') \phi (x') \phi (x).
\label{ttbar}
\eea

\subsection{ Generating Functional in Non-Equilibrium QED}

Extending the same analysis to QED we find the generating functional in non-equilibrium QED
\bea
&& Z[J_+,J_-,\eta_+,\eta_-,{\bar \eta}_+,{\bar \eta}_-]=\int [dQ_+] [dQ_-][d{\bar \psi}_+] [d{\bar \psi}_-] [d \psi_+ ] [d\psi_-]~ \times \nonumber \\
&& {\rm exp}[i\int d^4x [-\frac{1}{4}({F}_{\mu \nu}^2[Q_+]-{F}_{\mu \nu}^2[Q_-])-\frac{1}{2 \alpha} (
(\partial_\mu {Q_+}^{\mu })^2-(\partial_\mu {Q_-}^{\mu })^2)
+ {\bar \psi}_+ \dslash [Q_+] \psi_+  \nonumber \\
&& - {\bar \psi}_- \dslash [Q_-] \psi_-
+ J_+ \cdot Q_+  - J_- \cdot Q_- +{\bar \eta}_+ \psi_+ -{\bar \eta}_- \psi_-+ \eta_+  {\bar \psi}_+ - \eta_-  {\bar \psi}_- ]] \nonumber \\
&& ~\times <Q_+,\psi_+,{\bar \psi}_+,0|~\rho~|0,{\bar \psi}_-,\psi_-,Q_->.
\label{zfqinon}
\eea
where $Q_\mu$ is the photon field and $J_\mu$ is the corresponding source, $\psi$ is the electron field and ${\bar \eta}$ is the
corresponding source. $\rho$ is the initial density of state. The state $|Q_\pm,\psi_\pm,{\bar \psi}_\pm,0>$ is the quantum state
corresponding to the field configurations $Q_\mu({\vec x},t=t_{in}=0)$, $\psi({\vec x},t=t_{in}=0)$ and ${\bar \psi}({\vec x},t=t_{in}=0)$
respectively.
\section{ Factorization of Soft and Collinear Divergences in Non-Equilibrium QED }
The generating functional in the background field method of QED is given by eq. (\ref{zfqi}). Extending this to
non-equilibrium QED we find
\bea
&& Z[\rho,A,J_+,J_-,\eta_+,\eta_-,{\bar \eta}_+,{\bar \eta}_-]=\int [dQ_+] [dQ_-][d{\bar \psi}_+] [d{\bar \psi}_-] [d \psi_+ ] [d\psi_-]~ \times \nonumber \\
&& {\rm exp}[i\int d^4x [-\frac{1}{4}({F}_{\mu \nu}^2[Q_+]-{F}_{\mu \nu}^2[Q_-])-\frac{1}{2 \alpha} (
(\partial_\mu {Q_+}^{\mu })^2-(\partial_\mu {Q_-}^{\mu })^2)
+ {\bar \psi}_+ \dslash [Q_++A_+] \psi_+  \nonumber \\
&& - {\bar \psi}_- \dslash [Q_-+A_-] \psi_-
+ J_+ \cdot Q_+  - J_- \cdot Q_- +{\bar \eta}_+ \psi_+ -{\bar \eta}_- \psi_-+ \eta_+  {\bar \psi}_+ - \eta_-  {\bar \psi}_- ]] \nonumber \\
&& ~\times <Q_+,\psi_+,{\bar \psi}_+,0|~\rho~|0,{\bar \psi}_-,\psi_-,Q_->
\label{zfqinon1}
\eea
where $\rho$ is the initial density of states. The fermion fields and the sources transform as
\bea
&& \psi'_+ = U_+~ \psi_+,~~~~~~~~~{\bar \psi}'_+ = {\bar \psi}_+ ~U_+^{-1},~~~~~~~~~~~~~~\eta'_+ = U_+~ \eta_+,~~~~~~~~~{\bar \eta}'_+ =
{\bar \eta}_+ U^{-1}_+ \nonumber \\
&& \psi'_- = U_-~ \psi_-,~~~~~~~~~\bar \psi'_- = {\bar \psi}_- ~U_-^{-1},~~~~~~~~~~~~~~\eta'_- = U_-~ \eta_-,~~~~~~~~~{\bar \eta}'_- =
{\bar \eta}_-~ U^{-1}_-
\label{ftnn}
\eea
where
\bea
U_+=e^{-ie \omega_+(x)},~~~~~~~~~~~~~~~U_-=e^{-ie \omega_-(x)}.
\label{purepm}
\eea

Changing ${{ \psi}}_\pm \rightarrow { \psi}'_{\pm }$, ${{\bar \psi}}_\pm \rightarrow {\bar \psi}'_{\pm }$ but keeping $Q_\pm$ fixed
we find from eq. (\ref{zfqinon1})
\bea
&& Z[\rho,A,J_+,J_-,\eta_+,\eta_-,{\bar \eta}_+,{\bar \eta}_-]=\int [dQ_+] [dQ_-][d{\bar \psi}_+] [d{\bar \psi}_-] [d \psi_+ ] [d\psi_-]~ \times \nonumber \\
&& {\rm exp}[i\int d^4x [-\frac{1}{4}({F}_{\mu \nu}^2[Q_+]-{F}_{\mu \nu}^2[Q_-])-\frac{1}{2 \alpha} (
(\partial_\mu {Q_+}^{\mu })^2-(\partial_\mu {Q_-}^{\mu })^2)
+ {\bar \psi}_+ \dslash [Q_+] \psi_+ - {\bar \psi}_- \dslash [Q_-] \psi_- \nonumber \\
&&+ J_+ \cdot Q_+  - J_- \cdot Q_- +{\bar \eta}_+ \psi'_+ -{\bar \eta}_- \psi'_-+ \eta_+  {\bar \psi}'_+ - \eta_-  {\bar \psi}'_-]] \nonumber \\
&& ~\times <Q_+,\psi'_+,{\bar \psi}'_+,0|~\rho~|0,{\bar \psi}'_-,\psi'_-,Q_->.
\label{zfqinon2}
\eea

Note that the state $|0,{\bar \psi}'_-,\psi'_-,Q_->$ is at initial time $t=t_{in}=0$. We find
\bea
&& Z[\rho,A,J_+,J_-,\eta_+,\eta_-,{\bar \eta}_+,{\bar \eta}_-]=\int [dQ_+] [dQ_-][d{\bar \psi}_+] [d{\bar \psi}_-] [d \psi_+ ] [d\psi_-]~ \times \nonumber \\
&& {\rm exp}[i\int d^4x [-\frac{1}{4}({F}_{\mu \nu}^2[Q_+]-{F}_{\mu \nu}^2[Q_-])-\frac{1}{2 \alpha} (
(\partial_\mu {Q_+}^{\mu })^2-(\partial_\mu {Q_-}^{\mu })^2)
+ {\bar \psi}_+ \dslash [Q_+] \psi_+ - {\bar \psi}_- \dslash [Q_-] \psi_-  \nonumber \\
&&+ J_+ \cdot Q_+  - J_- \cdot Q_- +{\bar \eta}_+ \psi'_+ -{\bar \eta}_- \psi'_-+ \eta_+  {\bar \psi}'_+ - \eta_-  {\bar \psi}'_-]] \nonumber \\
&& ~\times <Q_+,\psi_+,{\bar \psi}_+,0|~\rho~|0,{\bar \psi}_-,\psi_-,Q_->
\label{zfqinon3}
\eea
which gives by using eq. (\ref{ftnn})
\bea
&& Z[\rho,A,J_+,J_-,\eta'_+,\eta'_-,{\bar \eta}'_+,{\bar \eta}'_-]=\int [dQ_+] [dQ_-][d{\bar \psi}_+] [d{\bar \psi}_-] [d \psi_+ ] [d\psi_-]~ \times \nonumber \\
&& {\rm exp}[i\int d^4x [-\frac{1}{4}({F}_{\mu \nu}^2[Q_+]-{F}_{\mu \nu}^2[Q_-])-\frac{1}{2 \alpha} (
(\partial_\mu {Q_+}^{\mu })^2-(\partial_\mu {Q_-}^{\mu })^2)
+ {\bar \psi}_+ \dslash [Q_+] \psi_+ - {\bar \psi}_- \dslash [Q_-] \psi_-  \nonumber \\
&&+ J_+ \cdot Q_+  - J_- \cdot Q_- +{\bar \eta}_+ \psi_+ -{\bar \eta}_- \psi_-+ \eta_+  {\bar \psi}_+ - \eta_-  {\bar \psi}_-]] \nonumber \\
&& ~\times <Q_+,\psi_+,{\bar \psi}_+,0|~\rho~|0,{\bar \psi}_-,\psi_-,Q_->
\label{zfqinon3}
\eea
Hence we find from eq. (\ref{zfqinon3})
\bea
Z[\rho,A,J_+,J_-,\eta'_+,\eta'_-,{\bar \eta}'_+,{\bar \eta}'_-]=Z[\rho,J_+,J_-,\eta_+,\eta_-,{\bar \eta}_+,{\bar \eta}_-]
\label{gn12}
\eea
in non-equilibrium QED in the presence of background field in pure gauge.

The correlation functions in non-equilibrium QED in the presence of background field $A_\mu(a)$ is given by
\bea
&&~\frac{\delta}{\delta {\bar \eta}_r(x_2) }~\frac{\delta}{\delta { \eta}_s(x_1) }~Z[\rho,A,J_+,J_-,\eta_+,\eta_-,{\bar \eta}_+,{\bar \eta}_-]~|_{J_+=J_-=\eta_+=\eta_-={\bar \eta}_+={\bar \eta}_-=0} \nonumber \\
&&~=~<\psi_r(x_2) {\bar \psi}_s(x_1)>_A~=~<in|\psi_r(x_2) {\bar \psi}_s(x_1)|in>_A \nonumber \\
\label{g1nnq}
\eea
where $r,s=+,-$ are the closed-time path indices. The correlation functions in non-equilibrium QED is given by
\bea
&& ~\frac{\delta}{\delta {\bar \eta}_r(x_2) }~\frac{\delta}{\delta { \eta}_s(x_1) }~Z[\rho,J_+,J_-,\eta_+,\eta_-,{\bar \eta}_+,{\bar \eta}_-]~|_{J_+=J_-=\eta_+=\eta_-={\bar \eta}_+={\bar \eta}_-=0} \nonumber \\
&&~=~<\psi_r(x_2) {\bar \psi}_s(x_1)>_{A=0}~=~<in|\psi_r(x_2) {\bar \psi}_s(x_1)|in>_{A=0}.~
\label{g1nnq0}
\eea
Hence we find from eqs. (\ref{gn12}), (\ref{g1nnq}) and (\ref{g1nnq0})
\bea
<in|\psi_r(x_2) {\bar \psi}_s(x_1)|in>_A~=~e^{-ie \omega_r(x_2)}~<in|\psi_r(x_2) {\bar \psi}_s(x_1)|in>_{A=0}~e^{ie \omega_s(x_1)}.
\label{fact1qnnq}
\eea
where all the $A_\mu(x)$ dependence has been factored into $e^{-ie \omega_r(x_2)}$ and $e^{ie \omega_s(x_1)}$.
Using eqs. (\ref{om3}) and (\ref{wilf1}) we find
\bea
&&~<in|\psi_r(x_2) {\bar \psi}_s(x_1)|in>_A~=~
{\rm exp}[-ie\int_{-\infty}^0 d\lambda~ h \cdot A_r(x_2^\mu +h^\mu \lambda )] ~\times \nonumber \\
&& <in|\psi_r(x_2) {\bar \psi}_s(x_1)|in>_{A=0}~ \times ~
{\rm exp}[-ie\int_0^{-\infty} d\lambda~ h \cdot A_s(x_1^\nu +h^\nu \lambda )]
\label{fact2tn}
\eea
which proves factorization of soft and collinear divergences in non-equilibrium QED.
This is similar to eq. (\ref{fact2t}) except that the closed-time path indices $r,s=+,-$ have appeared in
eq. (\ref{fact2tn}). Note that the repeated indices $r$ and $s$ in eq. (\ref{fact2tn}) are not summed.
\section{ Conclusions }
In order to study hadron production from quark-gluon plasma using the
non-equilibrium fragmentation function \cite{nayakfrag} we will need to prove factorization in
non-equilibrium QCD. Factorization refers to separation
of short distance from long distance effects in field theory. However, before proving factorization in
non-equilibrium QCD we need to show a similar factorization in non-equilibrium QED. In this paper we have proved
factorization of soft and collinear divergences in non-equilibrium QED by using Schwinger-Keldysh closed-time path
integral formalism in the background field method of QED in pure gauge.

The proof of factorization of fragmentation function in non-equilibrium
QCD at RHIC and LHC will be addressed elsewhere. This may be relevant to study hadron
production \cite{nayakfrag} from quark-gluon plasma \cite{qgp} at RHIC and LHC.
\acknowledgements

This work was supported in part by Department of Energy under contracts DE-FG02-91ER40664,
DE-FG02-04ER41319 and DE-FG02-04ER41298.

\end{document}